# Deterministic Many-Resonator *W* Entanglement of Nearly Arbitrary Microwave States via Attractive Bose-Hubbard Simulation

A. A. Gangat,* I. P. McCulloch, and G. J. Milburn

*ARC Centre for Engineered Quantum Systems, School of Mathematics and Physics, The University of Queensland, St. Lucia, Queensland 4072, Australia*

Multipartite entanglement of large numbers of physically distinct linear resonators is of both fundamental and applied interest, but there have been no feasible proposals to date for achieving it. At the same time, the Bose-Hubbard model with attractive interactions (ABH) is theoretically known to have a phase transition from the superfluid phase to a highly entangled nonlocal superposition, but observation of this phase transition has remained out of experimental reach. In this theoretical work, we jointly address these two problems by (1) proposing an experimentally accessible quantum simulation of the ABH phase transition in an array of tunably coupled superconducting circuit microwave resonators and (2) incorporating the simulation into a highly scalable protocol that takes as input any microwave-resonator state with negligible occupation of number states $|0\rangle$ and $|1\rangle$ and nonlocally superposes it across the whole array of resonators. The large-scale multipartite entanglement produced by the protocol is of the *W* type, which is well known for its robustness. The protocol utilizes the ABH phase transition to generate the multipartite entanglement of all of the resonators in parallel, and is therefore deterministic and permits an increase in resonator number without any increase in protocol complexity; the number of resonators is limited instead by system characteristics such as resonator-frequency disorder and inter-resonator coupling strength. Only one local and two global controls are required for the protocol. We numerically demonstrate the protocol with realistic system parameters and estimate that current experimental capabilities can realize the protocol with high fidelity for greater than 40 resonators. Because superconducting-circuit microwave resonators are capable of interfacing with other devices and platforms such as mechanical resonators and (potentially) optical fields, this proposal provides a route toward large-scale *W*-type entanglement in those systems as well.

           Subject Areas: Condensed Matter Physics, Quantum Physics, Quantum Information

## I. INTRODUCTION

Entanglement is an essential resource for a wide range of fundamental and applied uses. The generation of entanglement among linear resonators, however, is a difficult problem, for the fundamental reason that nonlinear processes are required to generate nonclassical states. Entanglement generation in linear resonators therefore requires the assistance of other nonlinear systems such as atoms or qubits, which makes the entanglement of physically distinct resonators a significant challenge. In 2011, Wang *et al.* [1] met this challenge by demonstrating deterministic entanglement of photons across two separate on-chip superconducting resonators with the assistance of tunable phase qubits and an auxiliary resonator. The protocol that they employed is extendible, with an increase in complexity, to more than two resonators, but the increase in complexity makes the protocol unfeasible for the many-resonator regime. Feasible protocols to achieve the entanglement of a large number of physically distinct resonators remain unformulated to date.

In a separate vein, the standard Bose-Hubbard model [2] consists of repulsive on-site interactions that compete with intersite hopping to give rise to the well-known superfluid-to-Mott-insulator quantum phase transition. In the superfluid phase, all of the bosons occupy a single reciprocal mode of the lattice and are consequently delocalized, but the phase is coherent across the whole lattice. For integer ratios of boson number to total number of lattice sites, the insulating phase arises when the on-site interaction energy is sufficiently larger than the hopping energy, so that the bosons are localized to single sites and number fluctuations on each site are suppressed to zero. The superfluid-to-Mott-insulator phase transition was observed in the landmark experiment of Ref. [3]. In contrast, the attractive Bose-Hubbard model (ABH) [4–8] has attractive on-site interactions and supports a (quasi-) [9] quantum phase transition wherein, as the attractive interactions increasingly dominate over the hopping, the ground state $|\Psi_{\text{gs}}\rangle$ changes from the superfluid phase to a phase in which the bosons are collectively confined to the same site but nonlocally superposed over all sites and number fluctuations at each site are amplified: $|\Psi_{gs}\rangle \approx \frac{1}{\sqrt{M}} \times (|N\rangle_1|0\rangle_2|0\rangle_3...|0\rangle_M + |0\rangle_1|N\rangle_2|0\rangle_3...|0\rangle_M + |0\rangle_1|0\rangle_2|N\rangle_3...|0\rangle_M + \cdots + |0\rangle_1|0\rangle_2|0\rangle_3...|N\rangle_M)$, where $M$ is the number of sites and $N$ is the number of bosons. As the ratio of hopping energy to interaction energy goes to zero, the ground state becomes exactly a $W$ state. The ABH phase transition has been theoretically considered within the context of cold atoms [4,5], trapped ions [8], and polaritons in cavity arrays [10], but experimental demonstration in these platforms remains a difficult and unmet challenge. Meanwhile, on-chip superconducting-circuit systems have emerged as a very effective platform for quantum electrodynamics [11–13]. Theoretical and experimental activity regarding such circuit QED systems has recently begun to move toward the many-body regime, where quantum coherence can potentially be achieved over 1000 or more interconnected microwave resonators [13]. Further, this platform allows the unique capability of *in situ* tunable coupling between resonators [14–17]. Access to the many-body regime and tunable coupling makes circuit QED an attractive option for quantum simulation of many-body Hamiltonians [13,18], but a proposal for the simulation of the ABH phase transition has not yet been formulated in this platform.

---
*a.gangat@physics.uq.edu.au

In this theoretical work, we show that recent experimental advances in superconducting circuits can be used to realize a circuit QED system wherein the ABH phase transition may be simulated in an array of tunably coupled superconducting microwave resonators. Further, we present a protocol that uses the ABH phase transition to convert almost any input state $|\psi_{in}\rangle$ of a single microwave resonator into a $W$-type state $\frac{1}{\sqrt{M}}(|\psi_{in}\rangle_1|0\rangle_2|0\rangle_3...|0\rangle_M + |0\rangle_1|\psi_{in}\rangle_2|0\rangle_3...|0\rangle_M + |0\rangle_1|0\rangle_2|\psi_{in}\rangle_3...|0\rangle_M + \cdots + |0\rangle_1|0\rangle_2|0\rangle_3...|\psi_{in}\rangle_M)$ that spans all $M$ resonators of the array, thereby deterministically generating discrete-variable multipartite entanglement of many resonators. The only fundamental restriction on $|\psi_{in}\rangle$ in our protocol is that it must have negligible occupation of Fock states $|0\rangle$ and $|1\rangle$. The scalability of our protocol to arrays with large $M$ is due to the fact that the ABH phase transition entangles all of the resonators simultaneously, rather than one by one, so that only one local and two global controls are required for the protocol, regardless of the number of resonators. We estimate that existing technology makes the protocol feasible for up to $M \approx 40$ resonators. Our protocol complements well the experimental capability demonstrated in Ref. [19] of producing arbitrary microwave states $|\psi_{in}\rangle$ in single on-chip microwave resonators.

The ability to deterministically create many-resonator $W$ entanglement of nearly arbitrary microwave states in superconducting circuits is of significance for many reasons. For instance, the qubit $W$ state $\frac{1}{\sqrt{M}}(|1\rangle_1|0\rangle_2|0\rangle_3...|0\rangle_M + |0\rangle_1|1\rangle_2|0\rangle_3...|0\rangle_M + |0\rangle_1|0\rangle_2|1\rangle_3...|0\rangle_M + \cdots + |0\rangle_1|0\rangle_2|0\rangle_3...|1\rangle_M)$ is known for the robustness of its entanglement under loss [20,21]. In particular, the global entanglement decay of such a state under both phase and amplitude damping is known to be independent of the number of qubits [21]. The $W$-type states that are output by our protocol (given above) may also be considered as qubit $W$ states because the $|\psi_{in}\rangle$ are always orthogonal to $|0\rangle$ because of the restriction on $|\psi_{in}\rangle$ mentioned above, and they therefore have the same robustness of entanglement. Therefore, although the mapping of the single-resonator state to a $W$ state does not increase the lifetime of the state itself, the entanglement that is generated by the process is well suited to the many-resonator regime. Further, it is of particular use for tests of nonlocality that utilize large-scale $W$ states [22,23]. Another significant aspect of our protocol is that when $|\psi_{in}\rangle$ are coherent states (with a sufficiently large amplitude), the protocol outputs entangled coherent states (ECSs) [24,25], which are of both fundamental and applied significance in their own right [25–29]. For example, coherent states of resonators are quasiclassical states, and many-resonator ECSs therefore constitute large-scale Schrödinger cats. To our knowledge, many-partite ECS generation has not been feasibly considered in any platform. Finally, microwave resonators in superconducting circuits may interface with other types of systems [30], such as various types of circuit and noncircuit qubits [30], mechanical resonators [31,32], and (potentially) optical fields [33–40]. The protocol presented here may therefore provide an indirect route toward large-scale $W$ entanglement in those systems that may be more feasible than other, more direct approaches.

## II. SYSTEM AND MODEL

As schematically illustrated in Fig. 1, we consider a one-dimensional lattice of microwave-frequency superconducting coplanar waveguide (CPW) resonators, each embedded with a SQUID that intersects the center conductor line, as conceptually introduced in Ref. [41]. The resonators have nearest-neighbor coupling via the tunable coupler demonstrated in Refs. [16,17]. As theoretically analyzed in Refs. [41,42], the SQUIDs can induce a negative Kerr nonlinearity in the microwave modes of each CPW such that, looking at only the fundamental mode $c$ of each resonator and considering uniform parameters across the lattice, the Hamiltonian of the system in the interaction picture is

$$\frac{H}{\hbar} = \sum_{j=1}^{M} -\frac{\chi}{2} c_j^\dagger c_j (c_j^\dagger c_j - 1) - \kappa(c_j^\dagger c_{j+1} + c_j c_{j+1}^\dagger), \quad (1)$$

where $\chi$ is positive, $\kappa$ may be positive or negative [16,17], $M$ is the total number of resonators in the lattice, $j$ is the resonator index, and periodic boundary conditions are assumed. Both $\chi$ and $\kappa$ may be tuned *in situ* via flux biases, and we designate their ranges as $0 \leq \chi \leq \chi_{max}$ and $-\kappa_{\max} \leq \kappa \leq \kappa_{\max}$. (We note that a similar circuit QED system with open boundary conditions and no tunable coupling is theoretically studied in the driven-dissipative regime in Ref. [43]. We also note that $\chi$ in such a setup may not be made strictly zero, but its minimal value may be roughly $2\pi \times 10$ kHz [41,42]. This residual value of $\chi$ may be canceled in each resonator through dispersive coupling of a qubit, as in Ref. [44], which demonstrated a positive Kerr nonlinearity of roughly $2\pi \times 1$ MHz.) Experimental considerations related to damping, higher-order terms of the nonlinearity, and resonator-frequency disorder are discussed in Sec. V.

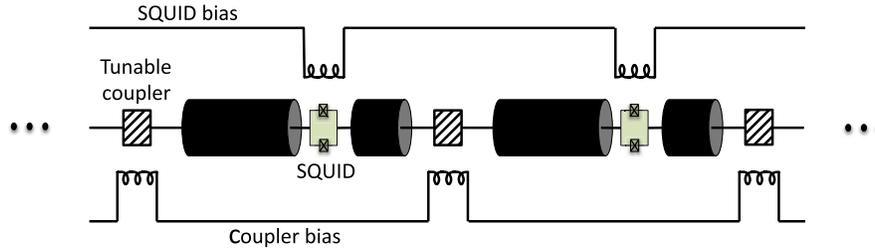

FIG. 1. System schematic. For simplicity, ground connections are not shown. CPW resonators (black shaded cylinders) form a one-dimensional array. Each resonator has a SQUID (green shaded boxes) connected to its center conductor and is coupled to its nearest neighbors with flux-tunable couplers (line-hatched boxes). The SQUIDs induce a flux-tunable nonlinearity into the resonator modes. Two bias lines are used to globally tune the SQUIDs and couplers separately. The first resonator (not shown) has a separate bias line for its SQUID.

Equation (1) is precisely the ABH model that is theoretically studied in Refs. [4–8]. For a fixed total number of quanta $N$ and $N > 1$, the ABH phase changes qualitatively as a function of the parameter $\tau = \frac{|\kappa|}{\chi(N-1)}$, as depicted in Fig. 2. $\tau_1 \approx 0.25$ is a characteristic constant of the ABH that is largely independent of lattice size [5,7]. When $\tau < \tau_1$, the attractive on-site energy dominates the hopping energy and the quanta in the eigenstates are collectively localized to single sites but superposed across all sites to form the nearly degenerate $W$ states

$$|\Psi_W^{(k)}\rangle = \frac{1}{\sqrt{M}} \sum_{j=1}^{M} e^{ik(2\pi j/M)} |N\rangle_j \prod_{r \neq j} |0\rangle_r, \quad (2)$$

where $k$ is an integer in the range $0 \leq k \leq M - 1$. These $W$ states become the exact eigenstates as $\tau \to 0$, and our proposal for ABH simulation is the only one to date that is able to access this perfect $W$-state regime, as it is the only one in which $\kappa$ is tunable to zero so that $\tau = 0$. $\tau_2$ is another characteristic constant of the ABH, such that for $\tau > \tau_2$, the ground state is a superfluid state. It is always the case that $\tau_1 \leq \tau_2$. When $\tau_1 \leq \tau \leq \tau_2$, the ABH phase is intermediate in nature between the $W$-state phase and the superfluid phase. $\tau_2$ increases monotonically with $M$: $\tau_2 = \tau_1$ for $M = 2$ [7], $\tau_1 < \tau_2 \lesssim 0.3$ for $3 \leq M \leq 5$ [6], and

$$\tau_2 \approx [2M\sin^2(\pi/M)]^{-1} \quad (3)$$

for $M \geq 5$ [5], which is approximately linear in $M$ and is plotted in Fig. 3. The approximate phase diagram for the ABH is represented in Fig. 4 with the regime $\tau_1 > \tau > \tau_2$ omitted.

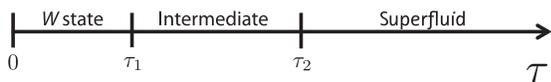

FIG. 2. Phases of the ABH as a function of the parameter $\tau$ ($\tau = \frac{|\kappa|}{\chi(N-1)}$). $\tau_1 \approx 0.25$ is approximately independent of lattice size, while $\tau_2$ ($\tau_2 \geq \tau_1$) is a function of $M$, as delineated in the text and plotted in Fig. 3.

We also consider the case when only the first site in the lattice may have a nonzero attractive interaction with strength $\chi_1$ and range $0 \leq \chi_1 \leq \chi_{\max}$:

$$\frac{H}{\hbar} = -\frac{\chi_1}{2} c_1^\dagger c_1 (c_1^\dagger c_1 - 1) - \sum_{j=1}^{M} \kappa(c_j^\dagger c_{j+1} + c_j c_{j+1}^\dagger). \quad (4)$$

For $N > 1$, the two parameter regimes of interest are $\chi_1(N-1)/2 \gg |\kappa|$, in which the quanta in the lowest-energy eigenstate are localized only at the first site, and $\chi_1(N-1)/2 \ll |\kappa|$, in which the quanta in the lowest-energy eigenstate form a superfluid state across the whole lattice for any finite value of $\kappa$. The approximate phase diagram for this model is represented in Fig. 5.

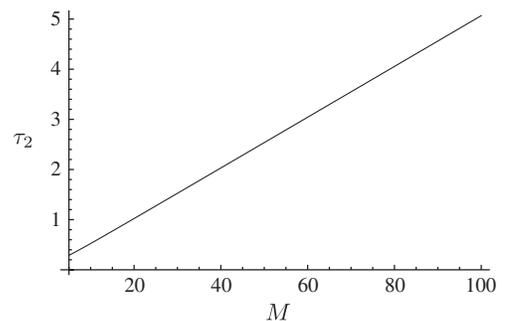

FIG. 3. A plot of Eq. (3) showing the dependence of the ABH parameter $\tau_2$ on the total number of lattice sites $M$ for $M \geq 5$.

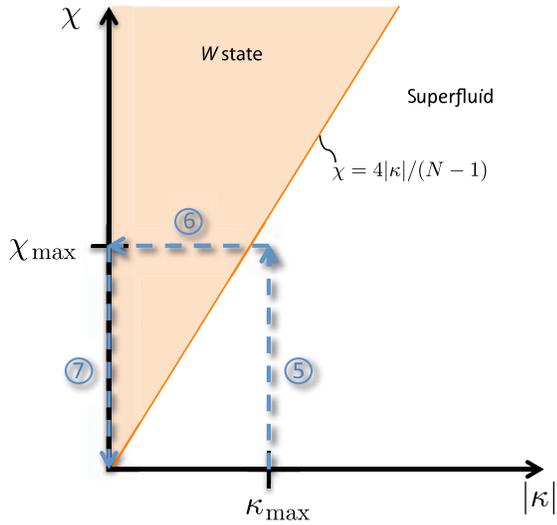

FIG. 4. Approximate phase diagram corresponding to the ABH of Eq. (1). The intermediate phase is not shown. Dashed blue arrows indicate the last three steps of the entanglement protocol.

Through local control of the nonlinearity $\chi_1$ of the first resonator and global control of $\chi_{j\neq 1}$ and $\kappa$, our proposed system can simulate both Eqs. (1) and (4). An initial state of the proposed system with $N \geq 2$ in the lowest-energy eigenstate and zero quanta in the other eigenstates may therefore adiabatically transition between the single-site localized phase, the delocalized phase, and the $W$ state. If instead the system starts in a superposition of different $N$ in the lowest-energy eigenstate, the linearity of quantum mechanics dictates that each component of the superposition with $N \geq 2$ may independently undergo the phase transitions. The entanglement protocol revolves around these phase transitions and therefore necessarily excludes initial states with number distributions that have a significant occupation of $N < 2$.

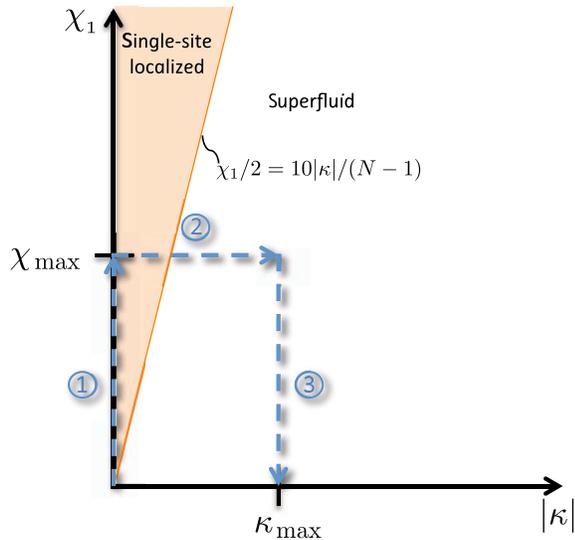

FIG. 5. Approximate phase diagram corresponding to the single-site nonlinearity model of Eq. (4). Dashed blue arrows indicate the first three steps of the entanglement protocol.

## III. ENTANGLEMENT PROTOCOL

Mention is made in the literature of using adiabatic transitions of Hamiltonians to generate entangled Fock states [10,45,46]. In particular, Hartmann et al. [10] suggest using an atom-cavity realization of the ABH to create a polaritonic (photon-atom hybrid) approximate $W$ state of single Fock states $|n\rangle$ (with $n > 1$) in multiple optical cavities via adiabatic transition from the superfluid regime ($\tau > \tau_2$) to the $W$-state regime ($\tau < \tau_1$) by tuning $\chi$. Here, we show that the ABH implemented in our proposed superconducting circuit can be used to create perfect $W$-type entanglement of arbitrary superpositions of photonic Fock states $|n\rangle$ ($n > 1$).

With the system in the vacuum state and $\chi = \kappa = 0$, the first resonator of the system is initialized in a state $|\psi_{\rm in}\rangle = \sum_n C_n |n\rangle$, where $|n\rangle$ denotes a Fock state with $n$ quanta and $C_n$ are complex amplitudes, so that the initial total system state is

$$|\Psi(0)\rangle = |\psi_{\rm in}\rangle_{j=1} \prod_{j=2}^{M} \otimes |0\rangle_j. \quad (5)$$

The probability distribution of $|\psi_{\rm in}\rangle$ in the number basis is therefore the probability distribution of the total number of quanta $N$ in the system. $|\psi_{\rm in}\rangle$ therefore has the constraint that $|C_n| \approx 0$ for $n < 2$. As mentioned at the end of Sec. II, each total quantum number $n$ from the distribution of $N$ may be treated independently. We may therefore express the total system state at all times as

$$|\Psi(t)\rangle = \sum_{n=0}^{\infty} C_n |\Psi_n(t)\rangle, \quad (6)$$

where $|\Psi_n(t)\rangle$ denotes the system state at time $t$ with definite total quantum number $n$, and the evolution of each $|\Psi_n(t)\rangle$ may be considered separately. The entanglement protocol employs manipulations of $\chi_1$, $\chi$, and $\kappa$ that simultaneously evolve each $|\Psi_n(t)\rangle$ from $|\Psi_n(0)\rangle = |n\rangle_{j=1}$ (a Fock state of the first resonator) into $|\Psi_n(T)\rangle = \frac{1}{\sqrt{M}} \times \sum_{j=1}^{M} |n\rangle_j \prod_{r \neq j} |0\rangle_r$ (a $W$ state spanning all of the resonators) for all $n > 1$. The total system state at the end of the protocol is therefore a $W$-type distribution of the input state $|\psi_{\rm in}\rangle$:

$$|\Psi(T)\rangle = \frac{1}{\sqrt{M}} \sum_{j=1}^{M} |\psi_{\rm in}\rangle_j \prod_{r \neq j} \otimes |0\rangle_r. \quad (7)$$

Because the entanglement protocol maintains each $|\Psi_n(t)\rangle$ in a single eigenstate of the system, we may express the time evolution as

$$|\Psi_n(t)\rangle = e^{-i\phi_n(t)} |n(t)\rangle_{\rm eig}, \quad (8)$$

where $\phi_n(t) = \frac{1}{\hbar}\int_0^t E_n(t')dt'$, and $E_n(t)$ denotes the energy of the eigenstate $|n(t)\rangle_{\text{eig}}$ that is occupied by the $n$ quanta at time $t$. We first describe the trajectory of $|n(t)\rangle_{\text{eig}}$ during the entanglement protocol to explain how it evolves from $|n(0)\rangle_{\text{eig}} = |n\rangle_{j=1}$ to $|n(T)\rangle_{\text{eig}} = \frac{1}{\sqrt{M}}\sum_{j=1}^{M}|n\rangle_j\prod_{r\neq j}|0\rangle_r$, and then explain how $e^{-i\phi_n(T)} = 1$ is achieved so that $|\Psi_n(T)\rangle = \frac{1}{\sqrt{M}}\sum_{j=1}^{M}|n\rangle_j\prod_{r\neq j}|0\rangle_r$. In general, $\phi_n(t) \neq \phi_m(t)$ when $n \neq m$, which can result in a distortion of the phase information contained in the initial state.

The steps of the entanglement protocol are illustrated with numbered arrows in the phase diagrams of Figs. 4 and 5 and in the pictorial representation of Fig. 6, which depicts the evolution of the full system state $|\Psi(t)\rangle$ [see Eq. (6)].

Let $t_x$ denote the time in the entanglement protocol after step number $x$. The first three steps of the protocol (see Figs. 5 and 6) use the model of Eq. (4) to convert each $|n\rangle$ localized in the first resonator into a delocalized superfluid state across the whole lattice. In the first step, $\chi_1$ is rapidly increased to $\chi_{\text{max}}$ in a time $\Delta t_1$, which does not change the occupied eigenstate: $|n(t_1)\rangle_{\text{eig}} = |n(0)\rangle_{\text{eig}}$. The attractive interaction thus introduced into the first resonator provides an energy barrier against which $\kappa$ may be tuned adiabatically in step 2 from 0 to $-\kappa_{\text{max}}$ in a time $\Delta t_2 \gg 4\kappa_{\text{max}}/[\chi_{\text{max}}(n-1)]^2$. In step 3, $\chi_1$ is tuned to zero (adiabatically with respect to the hopping dynamics that redistributes the quanta in real space) in a time $\Delta t_3 \gg \chi_{\text{max}}(n-1)/2\kappa_{\text{max}}^2$. Because of the negative sign of $\kappa$ and the single-site nonlinearity during the first three steps, $|n(t_3)\rangle_{\text{eig}}$ is actually the highest-energy eigenstate of the system. In step 4, $\kappa$ is rapidly tuned from $-\kappa_{\max}$ to $+\kappa_{\max}$ in a time $\Delta t_4$ so that $|n(t_4)\rangle_{\text{eig}}$ is the ground state of the system. $|n(t_4)\rangle_{\text{eig}}$ can therefore adiabatically connect to the lowest-energy $W$ state $|\Psi_W^{(k=0)}\rangle$ of Eq. (2). Steps 5 and 6 of the protocol (see Figs. 4 and 6) achieve this adiabatic connection by using the ABH [Eq. (1)]. The adiabaticity of these steps is with respect to two different physical processes: In the region $\tau/\tau_2 > 1$, Ref. [5] shows that the distribution of the quanta among the reciprocal modes is roughly constant with $\tau$, and the adiabaticity must therefore be with respect to the hopping dynamics that changes the intersite phase relationships; in the region $0 < \tau/\tau_2 < 1$, Ref. [5] shows that the reciprocal-space quantum distribution changes approximately linearly with $\tau$, and the adiabaticity must therefore be with respect to the on-site interaction, which redistributes the quanta in the reciprocal basis. (Analytical details are given in Sec. V B.) In step 5, the global parameter $\chi$ is adiabatically tuned from zero to $\chi_{\max}$ in a time $\Delta t_5$, so that $\tau = \kappa_{\max}/\chi_{\max}(n-1)$. Step 5 may place $|n(t_5)\rangle_{\text{eig}}$ in any of the three phases of the ABH (see Fig. 2), depending on the exact values of $\chi_{\max}$, $\kappa_{\max}$, and $n$. In step 6, $\kappa$ is adiabatically tuned from $\kappa_{\max}$ to zero in a time $\Delta t_6$, so that $\tau = 0$ and $|n(t_6)\rangle_{\text{eig}}$ is the $W$ state $|\Psi_W^{(k=0)}\rangle$. In this step, although the eigenstates become degenerate as $\tau \to 0$, the on-site attractive interaction energy proportional to $\chi$ serves as an energy barrier that protects the system from transitioning out of the ground state as long as the tuning is adiabatic with respect to $\chi$. As $\kappa$ is now off, the superposition is locked into place and $\chi$ may be rapidly tuned in step 7 from $\chi_{\max}$ to zero in a time $\Delta t_7$ without altering the occupied eigenstate: $|n(T)\rangle_{\text{eig}} = \frac{1}{\sqrt{M}} \sum_{j=1}^M |n\rangle_j \prod_{r \neq j} |0\rangle_r$. As the nonlinearity is now off, the CPWs are not hybridized with the SQUIDs and the state of the system is purely photonic.

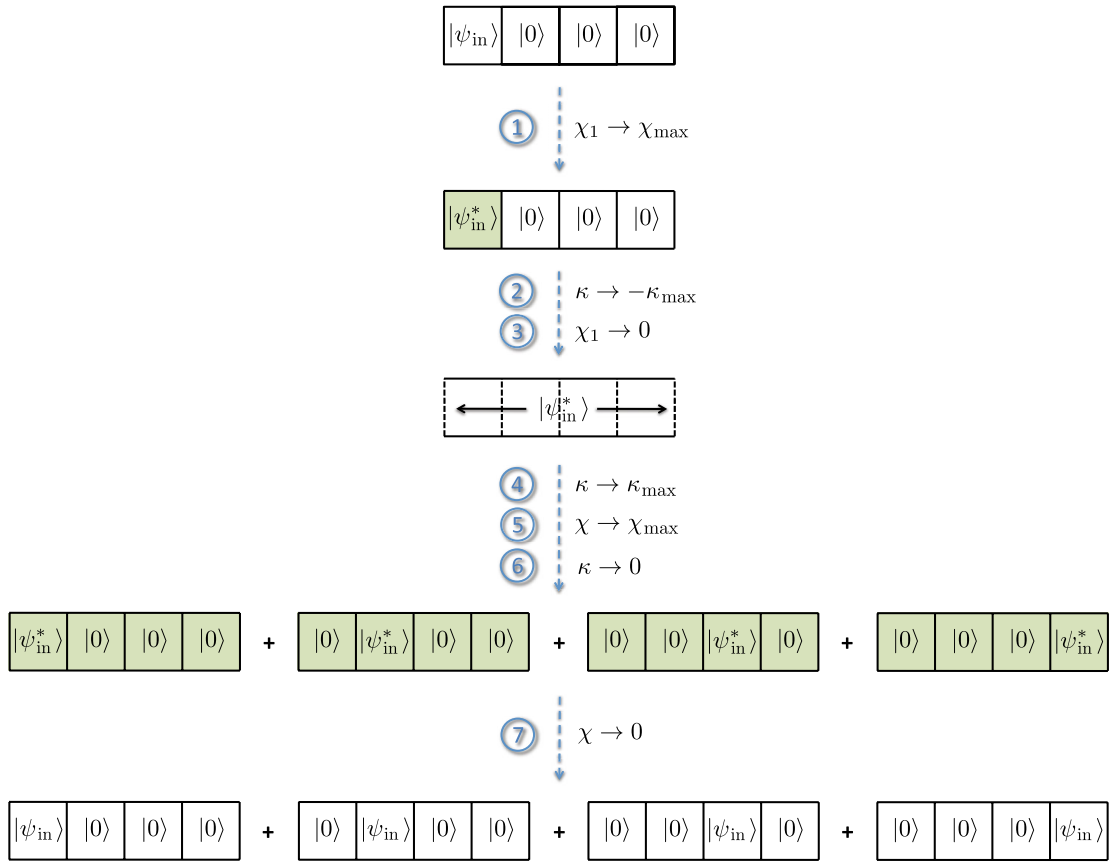

FIG. 6. Illustration of the entanglement protocol for the case of four resonators. Square boxes denote single resonators, solid black lines shared between square boxes denote inter-resonator coupling ($\kappa$) off, dashed black lines denote coupling on, green shading denotes resonator nonlinearity ($\chi$) on, white shading denotes nonlinearity off, periodic boundary conditions are implied, and dashed blue arrows denote steps of the entanglement protocol. $|\psi_{\text{in}}^*\rangle = \sum_n e^{-i\phi_n(t)}C_n|n\rangle$ denotes a version of $|\psi_{\text{in}}\rangle$ with modifications to the phases of each number component. A detailed account of the process in terms of the individual $|n\rangle$ is provided in the main text. In the first step, the nonlinearity is turned on for the first site only, which creates an energy barrier against which $\kappa$ may be turned on adiabatically in step 2. $|\psi_{\text{in}}^*\rangle$ may then adiabatically transition to the superfluid state by adiabatically turning off $\chi_1$ (step 3). At this point, $|\psi_{\text{in}}^*\rangle$ is in the highest excited superfluid state because of the lattice asymmetry introduced in step 1 and the sign of $\kappa$. After switching the sign of $\kappa$ in step 4, $|\psi_{\text{in}}^*\rangle$ is in the superfluid ground state. Step 5 ($\chi$ is adiabatically turned on across the whole lattice) and step 6 ($\kappa$ is adiabatically turned off) induce the ABH phase transition and make $\tau = 0$, so that the perfect $W$ state is achieved, with $|\psi_{\text{in}}^*\rangle$ nonlocally superposed in each resonator. Finally, turning off the resonator nonlinearities $\chi$ (step 7) restores the lattice to an array of uncoupled linear resonators that now contains the entangled state. With correct timing of the last step, $\phi_n(T) = 0$ ($T$ is the total time of the protocol) and $|\psi_{\text{in}}^*\rangle$ is restored to $|\psi_{\text{in}}\rangle$.

The effects of the entanglement protocol on $\phi_n(T)$ may be understood by separately considering the contributions of the on-site nonlinearity terms and the nearest-neighbor hopping terms of Eqs. (1) and (4). In the case of a single site, the nonlinearity contributes a number-dependent phase $n \int_0^t \frac{\chi_j(t')}{2} dt'$ [47], so that appropriate timing can make $\int_0^T \frac{\chi_j(t')}{2} dt'$ equal to an integer multiple of $2\pi$ and the effective contribution to $\phi_n(T)$ is zero for all $n$. It may be conjectured that with appropriate timing, the same type of cancellation for all $n$ could occur in the multisite case. However, this conjecture is far less trivial due to the changing distribution of the quanta with time. Remarkably, numerical investigations show the conjecture to be correct. For the hopping terms, it can also be conjectured that the contribution to $\phi_n(T)$ can be approximately canceled by simply making $\Delta t_2 = \Delta t_6$ and $\Delta t_3 = \Delta t_5$, so that the phase evolution due to the hopping terms when $\kappa < 0$ (steps 2 and 3) may cancel the phase evolution due to the hopping terms when $\kappa > 0$ (steps 5 and 6) because of the opposite signs of the hopping energy and because step 5 is in some sense the reverse of step 3 superposed on each site. Remarkably, numerical investigations reveal that this cancellation is also exact, even in the absence of adiabaticity. It is therefore possible to achieve $\phi_n(T) = 0$ through appropriate timing of the steps of the entanglement protocol, as we demonstrate in the numerical simulations below. As an alternative to the cancellation of $\phi_n(T)$ through timing, a calibration may be done whereby the phase $\phi_n(T)$ is measured after test runs of the protocol with $|\psi_{\text{in}}\rangle = |n\rangle_1$ for different $n$, so that the arbitrary input states generated by, for example, the method in Ref. [19] may be prepared with appropriate offset phases for each number component, so that the $\phi_n(T)$ accumulated through the protocol are all canceled.

## IV. DETECTION

Verification of $W$-state creation may be done by employing bipartite Wigner tomography [1] between different pairs of resonators to reconstruct their joint density matrix. This tomography is the technique that was used in Ref. [1] to show the creation of a NOON state of two superconducting-circuit resonators, and it should therefore be readily applicable to our proposed system. It does require, however, coherent drive access to each individual resonator that is to be measured as well as qubits coupled to each such resonator. Individual drive access should be possible, however, for proof-of-principle experiments in the few-resonator regime. The tomography entails a series of identical state preparations of the system, each one followed by destructive measurements of the two selected resonators by their corresponding qubits. A sufficient number of such measurements yields enough information to approximate the joint density matrix of the two resonators. By performing this process with different pairs of resonators after the entanglement protocol is run, the nature of the full system state after the entanglement protocol runs may be inferred. The technical details of the tomography process may be found in the Supplemental Material of Ref. [1].

Alternatively, if after the entangled state is prepared the system is made into a linear network by quenching $\kappa$ from zero to $\kappa_{\max}$, the protocol of Tufarelli *et al.* [48] allows for a single qubit tunably coupled to any resonator to be used to reconstruct the state of the entire array. This protocol may be more suitable for systems with larger numbers of resonators.

## V. EXPERIMENTAL CONSTRAINTS

With present-day devices [16,17,42], $\kappa_{\max}/2\pi$ and $\chi_{\max}/2\pi$ may reach as high as hundreds of MHz and may be tuned on a time scale of a few nanoseconds.

However, $\kappa$ should be limited to about $\kappa_{\max}/2\pi = 30$ MHz in the interest of preserving high experimental fidelity of the hopping Hamiltonian with the tunable coupler [49]. Modes $c_j$ have flux-dependent $T_1$ and $T_\phi$ because of the flux-dependent hybridization of the CPWs with the SQUIDs. $Q$ factors of over $1 \times 10^6$ have been demonstrated for on-chip CPWs [50], which gives $T_1 > 20\ \mu$s for the $2\pi \times 7.5$ GHz CPWs that we assume here. As mentioned in Ref. [42], the SQUIDs embedded in the CPW resonators should be able to achieve very long coherence times, considering the recent $T_1$ and $T_2$ measurements of the Josephson junction qubits in Refs. [51–53]. The 2D Xmon qubit demonstrated in Ref. [53], for example, shows a maximum $T_1$ of approximately 44 $\mu$s and $T_1 \approx T_\phi/2 \approx 20\ \mu$s at the flux-insensitive point. Further, considering an asymmetric SQUID as in Ref. [42] allows for $T_\phi \gtrsim 1$ ms at the flux value, where $\chi$ is maximum [54]. For our purposes here, we therefore approximate a constant $T_1$ value of 20 $\mu$s and a flux-dependent $T_\phi$ value that varies linearly from 1 s to 300 $\mu$s as $\chi$ is increased linearly from 0 to $\chi_{\max}$.

In this section, we show that these and other experimental capabilities make our protocol feasible for array sizes of greater than 40 resonators. In particular, for the choices of $\kappa_{\max}/2\pi = 30$ MHz and $\chi_{\max}/2\pi = 14$ MHz, the ranges $2 \leq n \leq 40$ and $2 < M \leq 42$ become available, which can accommodate high-amplitude many-partite ECSs.

### A. Higher-order terms of CPW nonlinearity

The full expression for the nonlinearity ($H_{\rm NL}$) introduced into the CPWs by the embedded SQUIDs is given in Eq. (27) of Ref. [42]. Looking only at the fundamental mode and using the relations $\Phi_0 = h/2e$, $E_J = (\frac{\Phi_0}{2\pi})^2/L_J$, $L_J = L'/\eta_l$, and $L' = 2E'_C/(\omega_{c,0}^2 e^2)$, we find

$$H_{\rm NL} = \sum_{i>1} \frac{(-1)^{i+1}}{2(2i!)}\left(\frac{4\chi_{\max}}{\omega_{c,m}}\right)^{i-1} \eta_l \hbar \omega_{c,m}[c^\dagger + c]^{2i}, \quad (9)$$

where $\eta_l$ ($0 \lesssim \eta_l \leq 1$) is the inductive participation ratio dependent upon the flux through the SQUID loop and $\omega_{c,m}$ is the frequency of the CPW mode $m$ in the absence of the SQUID. After the rotating-wave approximation, to be able to neglect nonlinear terms higher than $(c^\dagger c)^2$, we require

$$\frac{1}{12}\left(\frac{\chi_{\max}}{\omega_{c,m}}\right)6(n_{\max})^2 \gg \frac{1}{30}\left(\frac{\chi_{\max}}{\omega_{c,m}}\right)^2 20(n_{\max})^3 \quad (10)$$

or

$$n_{\max} \ll \frac{3}{4}\frac{\omega_{c,m}}{\chi_{\max}}. \quad (11)$$

Larger values of $\chi_{\max}$ come at a cost of reduced $n_{\max}$ but allow for quicker adiabatic tuning of $\kappa$, as per the discussion below. Smaller values of $\chi_{\max}$ enable larger array sizes, also as per below. Larger values of $n_{\max}$ may be accommodated for a fixed $\chi_{\max}$ by using higher-order CPW resonator modes. If we assume a fundamental mode frequency of $\omega_{c,1}/2\pi = 7.5$ GHz, selecting $\chi_{\max}/2\pi = 14$ MHz allows $n_{\max} = 40$ when the fundamental mode is used, selecting $\chi_{\max}/2\pi = 25$ MHz allows $n_{\max} = 23$ when the fundamental mode is used, and selecting $\chi_{\max}/2\pi = 100$ MHz allows $n_{\max} = 6$ when the fundamental mode is used.

### B. Adiabaticity

Figure 7 of Ref. [5] reveals a universal behavior of the ABH in which the population of the fundamental normal mode of the lattice decreases approximately linearly with the parameter $\tau/\tau_2$ in the region $0 \leq \tau/\tau_2 \leq 1$, and is approximately constant in the region $\tau/\tau_2 > 1$. This universal behavior translates to the two adiabatic constraints $\frac{d\tau/\tau_2}{dt} \ll \chi_{\max}$ and $\frac{d\tau/\tau_2}{dt} \ll 2\kappa_{\max}$ in the respective regions.

We denote the value of $\tau$ at the end of step 5 as $\tau^* = \frac{\kappa_{\max}}{\chi_{\max}(n-1)}$. The consideration in the next subsection shows that it is of interest to minimize the time taken to tune $\tau$ from $\tau_2$ to $\tau_1$. We find that the minimization occurs when $\tau^* \approx \tau_2$. (Although the minimization is not simultaneously possible for all $n$, it is sufficient to assume so as a rough estimate.) In this case, the adiabatic constraint to tune $\chi$ from 0 to $\chi_{\max}$ in a time $\Delta t_5$ is $\frac{d\chi}{dt} \lesssim \frac{1}{5}\chi^2\tau_2(n-1)$. Since the constraint only becomes applicable as $\tau$ approaches $\tau_2$, the approximation $\chi \to \chi''$ is made on the right-hand side to yield

$$\Delta t_5 \gtrsim 5/\kappa_{\max}. \quad (12)$$

For step 6, we have the requirement $\frac{d\kappa}{dt} \lesssim \frac{1}{10}\chi^2_{\max}(n-1)\tau_2$. Tuning $\kappa$ from $\kappa_{\max}$ to $\kappa(\tau_1)$ in a time $\Delta t_{6a}$ requires $\Delta t_{6a} \gtrsim 10\frac{\tau_2 - \tau_1}{\chi_{\max}\tau_2}$, and tuning $\kappa$ from $\kappa(\tau_1)$ to zero in a time $\Delta t_{6b}$ requires $\Delta t_{6b} \gtrsim 10\frac{\tau_1}{\chi_{\max}\tau_2}$. The constraint on step 6 is therefore

$$\Delta t_6 \gtrsim 10/\chi_{\max}. \quad (13)$$

For $\chi_{\max}/2\pi$ and $\kappa_{\max}/2\pi$ on the scale of tens of MHz, this constraint puts steps 5 and 6 on the 10–100 ns time scale, which is well below the estimated $T_1$ and $T_\phi$ of the resonators.

### C. Disorder

Disorder in inter-resonator coupling frequencies can be neglected when resonator frequencies greatly exceed the coupling frequencies [55], which is the case here. We focus therefore upon disorder in resonator frequencies.

The inevitable frequency spread $\Delta\omega$ of the resonators relates to the discussion in Ref. [4] concerning the fact that the nondegeneracy of the lattice sites induces nonuniformity in $|(n,t)\rangle_{\rm eig}$ in the site basis at a rate $1/\Delta\omega$. This undesirable dynamics is relevant in the intermediate regime $\tau_1 < \tau < \tau_2$, which is traversed in step 6, given the assumptions above. If this intermediate regime is traversed on a time scale comparable to or longer than $1/\Delta\omega$, the quanta have time to localize in the lowest-energy site. In order to achieve symmetric $W$ entanglement with high fidelity, we therefore require $\Delta t_{\text{int}} \ll 1/\Delta\omega$, where $\Delta t_{\text{int}}$ is the time it takes to traverse the intermediate regime. From the previous subsection, $\Delta t_{\text{int}} = \Delta t_{6a}$. Letting $\Delta t_{\text{int}} = 1/10\Delta\omega$, we find $\frac{1}{\Delta\omega} \gtrsim \frac{100}{\chi_{\max}}(1 - \tau_1/\tau_2)$, where it is assumed that $\tau_2 = \frac{\kappa_{\max}}{\chi_{\max}(n_{\min}-1)}$. This inequality gives

$$\Delta\omega \lesssim \frac{\chi_{\max}\kappa_{\max}}{100[\kappa_{\max} - \chi_{\max}(n_{\min}-1)/4]}, \quad (14)$$

which gives a lower bound on $\chi_{\max}$. Smaller values of $\chi_{\max}$ (or $n_{\min}$) enable larger $\tau_2$ (and therefore larger $M$) for a fixed $n_{\min}$ (or $\chi_{\max}$) but also require smaller values of $\Delta\omega$ and larger $\Delta t_6$. Current capabilities demonstrate $\Delta\omega \gtrsim 2\pi \times 1$ MHz for GHz-frequency CPW resonators [55]. Choosing $n_{\min} = 2$ and $\kappa_{\max}/2\pi = 30$ MHz, we thereby find the constraint $\chi_{\max}/2\pi \gtrsim 14$ MHz. Using Eqs. (3) and (11), this constraint translates to $M \leq 42$ and $n_{\max} \leq 40$, which is sufficient for many-partite ECS creation. If $\Delta\omega$ is reduced by a factor of 2, we find $\chi_{\max}/2\pi \gtrsim 7.5$ MHz and $M \leq 80$ for $n_{\min} = 2$. Reducing $\Delta\omega$ to $2\pi \times 0.1$ MHz gives $\chi_{\max}/2\pi \gtrsim 1.6$ MHz and $M \leq 360$ for $n_{\min} = 2$.

## VI. SIMULATIONS

We demonstrate the entanglement protocol by numerically integrating the master equation

$$\frac{d\rho}{dt} = -\frac{i}{\hbar}[H, \rho] + \sum_{j=1}^{M} \frac{1}{T_1}\mathcal{D}[c_j]\rho + \sum_{j=1}^{M} \frac{1}{T_\phi}\mathcal{G}[c_j]\rho, \quad (15)$$

where $\rho$ is the density matrix for the CPW chain, $H$ is the system Hamiltonian, $\mathcal{D}[c_j]\rho = c_j\rho c_j^\dagger - c_j^\dagger c_j\rho/2 - \rho c_j^\dagger c_j/2$ is the amplitude-damping operator, $\mathcal{G}[c_j]\rho = c_j^\dagger c_j\rho c_j^\dagger c_j - (c_j^\dagger c_j)^2\rho/2 - \rho(c_j^\dagger c_j)^2/2$ is the phase-damping operator, $1/T_1$ is the amplitude-damping rate, and $1/T_\phi$ is the phase-damping rate. The integration is done using the fourth-order Runge-Kutta method with a time step of size $10^{-12}$ s. As discussed in Sec. V, we model the damping with a constant $T_1$ value of 20 $\mu$s and a flux-dependent $T_\phi$ value that varies linearly from 1 s to 300 $\mu$s as $\chi$ is increased linearly from 0 to $\chi_{\max}$.

In the first simulation, a system of three resonators ($M = 3$) is considered with an input state $|\psi_{\text{in}}\rangle = \frac{1}{\sqrt{2}}|2\rangle + \frac{1}{\sqrt{2}}|3\rangle$. The system parameters $\chi_{\max}/2\pi = 100$ MHz and $\kappa_{\max}/2\pi = 30$ MHz are used, which satisfies the constraints discussed in Sec. V, and the timings of the protocol steps are chosen to respect the adiabaticity constraints. The Hilbert space at each site is truncated at $n = 3$. The fidelity $|\langle\Psi(t)|\Psi_W\rangle|^2$ of the system state $|\Psi(t)\rangle$ with the target state $|\Psi_W\rangle = \frac{1}{\sqrt{3}}\sum_{j=1}^{3}|\psi_{\text{in}}\rangle_j\prod_{r\neq j}\otimes|0\rangle_r$ during the final step (step 7) of the protocol is shown in Fig. 7. The total physical time of the simulation is $0.1064$ $\mu$s, at the end of which the fidelity is about 97.5%. The same simulation is also performed without damping (not shown), and a peak fidelity of 98.6% is found. Further numerical investigations (not shown) reveal that in the case of no damping, the peak fidelity is limited only by imperfect adiabaticity. Finally, the same simulation is also performed (not shown) without damping but with added disorder, such that the first site has a frequency that is $2\pi \times 0.5$ MHz ($2\pi \times 1$ MHz) higher than the second site, and the third site has a frequency that is $2\pi \times 0.5$ MHz ($2\pi \times 1$ MHz) lower than the second site. In this case, the fidelity peak corresponding to the first peak in Fig. 7 drops to about 96.8% (92.5%) and the second peak drops to about 94.7% (85.9%). As there is no damping, this discrepancy between the first and second fidelity peaks can be understood as being due to the disorder, which causes the phases to evolve at different rates on each side. This effect of the disorder indicates that disorder may place limits on the total time length of the protocol for cases where intersite phase differences would be undesirable.

In the second simulation, we use $M = 3$, $\chi_{\max}/2\pi = 120$ MHz, and $\kappa_{\max}/2\pi = 30$ MHz. The input state is $|\psi_{\text{in}}\rangle = \frac{1}{\sqrt{6}}|2\rangle + \frac{2}{\sqrt{6}}|3\rangle + e^{i\pi/9}\frac{1}{\sqrt{6}}|4\rangle$, and the Hilbert space at each site is truncated at $n=4$. The fidelity $|\langle\Psi(t)|\Psi_W\rangle|^2$ during step 7 is shown in Fig. 8. The timings of the protocol steps are based on trial simulations that determine the level of adiabaticity needed to achieve the high peak fidelity shown. The total physical time of the second simulation is $0.3556~\mu$s.

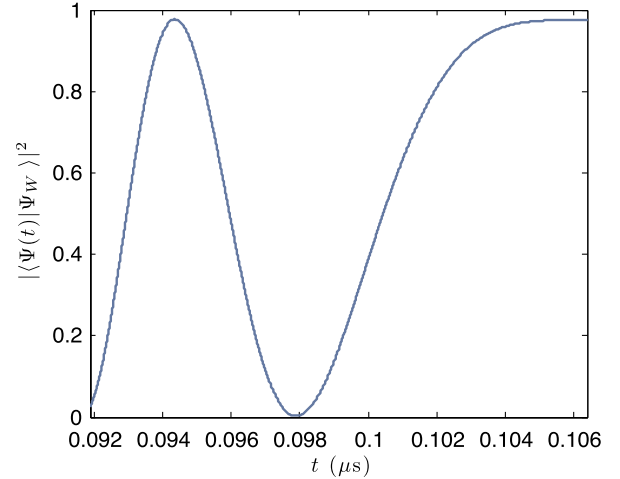

FIG. 7. Fidelity of the system state $|\Psi(t)\rangle$ with the target state $|\Psi_W\rangle = \frac{1}{\sqrt{3}}\sum_{j=1}^{3}|\psi_{\text{in}}\rangle_j\prod_{r\neq j}\otimes|0\rangle_r$ during the final step of the entanglement protocol with system parameters $\chi_{\max}/2\pi = 100$ MHz, $\kappa_{\max}/2\pi = 30$ MHz, and $M = 3$ and damping as explained in the text. The input state is $|\psi_{\text{in}}\rangle = \frac{1}{\sqrt{2}}|2\rangle + \frac{1}{\sqrt{2}}|3\rangle$. The fidelity oscillations are due to the phases $\phi_2(t)$ and $\phi_3(t)$ of the respective number components evolving at different rates because of the number-dependent frequency induced by the Kerr nonlinearity at each site. The frequency of the oscillations decays to zero as $\chi$ is tuned to zero, and the timing $\Delta t_7$ of the step is chosen such that the oscillations cease at a fidelity peak of about 97.5%.

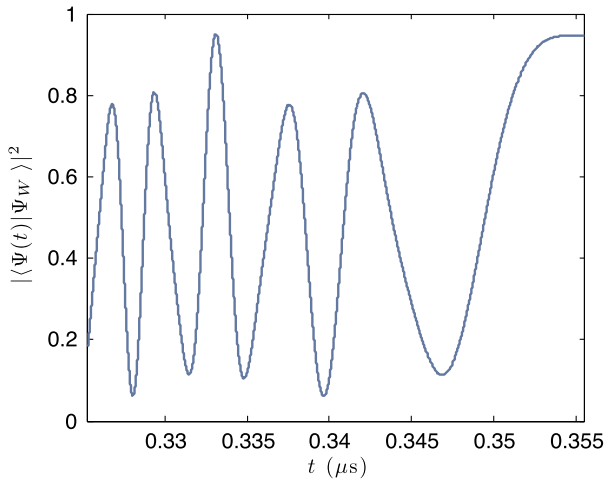

FIG. 8. Fidelity of the system state $|\Psi(t)\rangle$ with the target state $|\Psi_W\rangle = \frac{1}{\sqrt{3}}\sum_{j=1}^{3}|\psi_{\rm in}\rangle_j \prod_{r\neq j}\otimes |0\rangle_r$ during the final step of the entanglement protocol with system parameters $\chi_{\rm max}/2\pi = 120$ MHz, $\kappa_{\rm max}/2\pi = 30$ MHz, and $M=3$ and damping as explained in the text. The input state is $|\psi_{\rm in}\rangle = \frac{1}{\sqrt{6}}|2\rangle + \frac{2}{\sqrt{6}}|3\rangle + e^{i\pi/9}\frac{1}{\sqrt{6}}|4\rangle$. The fidelity oscillations are due to the phases of the number components evolving at different rates because of the number-dependent frequency induced by the Kerr nonlinearity at each site. Each period of the oscillations has three peaks because of the three frequency differences of the three number components of $|\psi_{\rm in}\rangle$. The frequency of the oscillations decays to zero as $\chi$ is tuned to zero, and the timing $\Delta t_7$ of the step is chosen such that the oscillations cease at a fidelity peak of about 95%.

## VII. CONCLUSIONS AND OUTLOOK

We have proposed a circuit QED simulation of the attractive Bose-Hubbard model in which the as-yet experimentally unobserved superfluid-to-$W$-state quantum phase transition may be realized with existing experimental capability. A unique aspect of this proposal is the capability of the *in situ* tuning of the hopping energy to enable access to the perfect $W$-state regime. We have further presented a protocol built around the attractive Bose-Hubbard simulation that deterministically produces $W$-type entanglement of nearly arbitrary single-resonator states over a large number of microwave resonators in parallel. We have numerically demonstrated our protocol with complex input states in an array of three resonators using realistic parameters and have shown the attainability of high output-state fidelity with the target state. The highly entangled large-scale states that the protocol is able to produce have both fundamental and applied significance.

Looking ahead, considering the equilibrium physics of the attractive Bose-Hubbard model in a circuit QED setup opens new prospects due to the unique flexibility of the platform. Studies of the ABH involving different types of lattice geometries, couplings, spatial modulations of parameters, defects, and controlled disorder that were not feasible in other platforms are a possibility in circuit QED. Further, it is interesting to consider the equilibrium and nonequilibrium phenomenologies that would result from interweaving the ABH and Jaynes-Cummings-Hubbard models in the same circuit. Also, the many-resonator entanglement protocol presented here for microwave resonators may offer a path toward creating large-scale $W$-type entanglement in mechanical and optical degrees of freedom because of the potential that superconducting circuits hold for interfacing with those platforms. Finally, it is worthwhile considering how lattice geometries and couplings different than the one-dimensional, nearest-neighbor coupling case considered here could allow the protocol to scale to larger numbers of resonators.

## ACKNOWLEDGMENTS

We acknowledge the support of the Australian Research Council Centre of Excellence for Engineered Quantum Systems (Grant No. CE110001013). A. A. G. acknowledges support from the University of Queensland and helpful discussions with Jerome Bourassa.

[1] H. Wang *et al.*, *Deterministic Entanglement of Photons in Two Superconducting Microwave Resonators*, Phys. Rev. Lett. **106**, 060401 (2011).

[2] M. P. A. Fisher, P. B. Weichman, G. Grinstein, and D. S. Fisher, *Boson Localization and the Superfluid-Insulator Transition*, Phys. Rev. B **40**, 546 (1989).

[3] M. Greiner, O. Mandel, T. Esslinger, T. W. Hänsch, and I. Bloch, *Quantum Phase Transition from a Superfluid to a Mott Insulator in a Gas of Ultracold Atoms*, Nature (London) **415**, 39 (2002).

[4] M. W. Jack and M. Yamashita, *Bose-Hubbard Model with Attractive Interactions*, Phys. Rev. A **71**, 023610 (2005).

[5] P. Buonsante, V. Penna, and A. Vezzani, *Attractive Ultracold Bosons in a Necklace Optical Lattice*, Phys. Rev. A **72**, 043620 (2005).

[6] P. Buonsante, P. Kevrekidis, V. Penna, and A. Vezzani, *Ground-State Properties of Attractive Bosons in Mesoscopic 1D Ring Lattices*, J. Phys. B **39**, S77 (2006).

[7] N. Oelkers and J. Links, *Ground-State Properties of the Attractive One-Dimensional Bose-Hubbard Model*, Phys. Rev. B **75**, 115119 (2007).

[8] X.-L. Deng, D. Porras, and J. I. Cirac, *Quantum Phases of Interacting Phonons in Ion Traps*, Phys. Rev. A **77**, 033403 (2008).

[9] The transition is not strictly a quantum phase transition as the thermodynamic limit for the ABH is ill defined [7].

[10] M. J. Hartmann, F. G. S. L. Brandão, and M. B. Plenio, *Strongly Interacting Polaritons in Coupled Arrays of Cavities*, Nat. Phys. **2,** 849 (2006).

[11] R. J. Schoelkopf and S. M. Girvin, *Wiring Up Quantum Systems*, Nature (London) **451,** 664 (2008).

[12] J. Q. You and F. Nori, *Atomic Physics and Quantum Optics Using Superconducting Circuits*, Nature (London) **474,** 589 (2011).

[13] A. A. Houck, H. E. Türeci, and J. Koch, *On-Chip Quantum Simulation with Superconducting Circuits*, Nat. Phys. **8,** 292 (2012).

[14] R. A. Pinto, A. N. Korotkov, M. R. Geller, V. S. Shumeiko, and J. M. Martinis, *Analysis of a Tunable Coupler for Superconducting Phase Qubits*, Phys. Rev. B **82,** 104522 (2010).

[15] B. Peropadre, D. Zueco, F. Wulschner, F. Deppe, A. Marx, R. Gross, and J. J. García-Ripoll, *Tunable Coupling Engineering between Superconducting Resonators: From Sidebands to Effective Gauge Fields*, Phys. Rev. B **87,** 134504 (2013).

[16] R. C. Bialczak *et al.*, *Fast Tunable Coupler for Superconducting Qubits*, Phys. Rev. Lett. **106,** 060501 (2011).

[17] Y. Yin *et al.*, *Catch and Release of Microwave Photon States*, Phys. Rev. Lett. **110,** 107001 (2013).

[18] S. Schmidt and J. Koch, *Circuit QED Lattices: Towards Quantum Simulation with Superconducting Circuits*, arXiv:1212.2070.

[19] M. Hofheinz *et al.*, *Synthesizing Arbitrary Quantum States in a Superconducting Resonator*, Nature (London) **459,** 546 (2009).

[20] H. J. Briegel and R. Raussendorf, *Persistent Entanglement in Arrays of Interacting Particles*, Phys. Rev. Lett. **86,** 910 (2001).

[21] R. Chaves and L. Davidovich, *Robustness of Entanglement as a Resource*, Phys. Rev. A **82,** 052308 (2010).

[22] L. Heaney, A. Cabello, M. França Santos, and V. Vedral, *Extreme Nonlocality with One Photon*, New J. Phys. **13,** 053054 (2011).

[23] Z. Wang and D. Markham, *Nonlocality and Entanglement for Symmetric States*, Phys. Rev. A **87,** 012104 (2013).

[24] B. C. Sanders, *Entangled Coherent States*, Phys. Rev. A **45,** 6811 (1992).

[25] B. C. Sanders, *Review of Entangled Coherent States*, J. Phys. A **45,** 244002 (2012).

[26] W. J. Munro, K. Nemoto, G. J. Milburn, and S. L. Braunstein, *Weak-Force Detection with Superposed Coherent States*, Phys. Rev. A **66,** 023819 (2002).

[27] J. Joo, W. J. Munro, and T. P. Spiller, *Quantum Metrology with Entangled Coherent States*, Phys. Rev. Lett. **107,** 083601 (2011).

[28] H. Jeong and N. B. An, *Greenberger-Horne-Zeilinger–Type and W-Type Entangled Coherent States: Generation and Bell-Type Inequality Tests without Photon Counting*, Phys. Rev. A **74,** 022104 (2006).

[29] N. B. An, *Optimal Processing of Quantum Information via W-Type Entangled Coherent States*, Phys. Rev. A **69,** 022315 (2004).

[30] Z.-L. Xiang, S. Ashhab, J. Q. You, and F. Nori, *Hybrid Quantum Circuits: Superconducting Circuits Interacting with Other Quantum Systems*, Rev. Mod. Phys. **85,** 623 (2013).

[31] C. A. Regal, J. D. Teufel, and K. W. Lehnert, *Measuring Nanomechanical Motion with a Microwave Cavity Interferometer*, Nat. Phys. **4,** 555 (2008).

[32] T. A. Palomaki, J. W. Harlow, J. D. Teufel, R. W. Simmonds, and K. W. Lehnert, *Coherent State Transfer between Itinerant Microwave Fields and a Mechanical Oscillator*, Nature (London) **495,** 210 (2013).

[33] Sh. Barzanjeh, M. Abdi, G. J. Milburn, P. Tombesi, and D. Vitali, *Reversible Optical-to-Microwave Quantum Interface*, Phys. Rev. Lett. **109,** 130503 (2012).

[34] Y.-D. Wang and A. A. Clerk, *Using Interference for High Fidelity Quantum State Transfer in Optomechanics*, Phys. Rev. Lett. **108,** 153603 (2012).

[35] Y.-D. Wang and A. A. Clerk, *Using Dark Modes for High-Fidelity Optomechanical Quantum State Transfer*, New J. Phys. **14,** 105010 (2012).

[36] L. Tian, *Adiabatic State Conversion and Pulse Transmission in Optomechanical Systems*, Phys. Rev. Lett. **108,** 153604 (2012).

[37] M. Hafezi, Z. Kim, S. L. Rolston, L. A. Orozco, B. L. Lev, and J. M. Taylor, *Atomic Interface between Microwave and Optical Photons*, Phys. Rev. A **85,** 020302(R) (2012).

[38] M. Tsang, *Cavity Quantum Electro-optics*, Phys. Rev. A **81,** 063837 (2010).

[39] M. Tsang, *Cavity Quantum Electro-optics. II. Input-Output Relations between Traveling Optical and Microwave Fields*, Phys. Rev. A **84,** 043845 (2011).

[40] M. Winger, T. Blasius, T. P. M. Alegre, A. H. Safavi-Naeini, S. Meenehan, J. Cohen, S. Stobbe, and O. Painter, *A Chip-Scale Integrated Cavity-Electro-optomechanics Platform*, Opt. Express **19,** 24905 (2011).

[41] M. Leib, F. Deppe, A. Marx, R. Gross, and M. J. Hartmann, *Networks of Nonlinear Superconducting Transmission Line Resonators*, New J. Phys. **14,** 075024 (2012).

[42] J. Bourassa, F. Beaudoin, J. M. Gambetta, and A. Blais, *Josephson-Junction-Embedded Transmission-Line Resonators: From Kerr Medium to In-Line Transmon*, Phys. Rev. A **86,** 013814 (2012).

[43] M. Leib and M. J. Hartmann, *Bose-Hubbard Dynamics of Polaritons in a Chain of Circuit Quantum Electrodynamics Cavities*, New J. Phys. **12,** 093031 (2010).

[44] A. J. Hoffman, S. J. Srinivasan, S. Schmidt, L. Spietz, J. Aumentado, H. E. Türeci, and A. A. Houck, *Dispersive Photon Blockade in a Superconducting Circuit*, Phys. Rev. Lett. **107,** 053602 (2011).

[45] J. I. Cirac, M. Lewenstein, K. Mølmer, and P. Zoller, *Quantum Superposition States of Bose-Einstein Condensates*, Phys. Rev. A **57,** 1208 (1998).

[46] C. Lee, *Adiabatic Mach-Zehnder Interferometry on a Quantized Bose-Josephson Junction*, Phys. Rev. Lett. **97,** 150402 (2006).

[47] D. F. Walls and G. J. Milburn, *Quantum Optics* (Springer-Verlag, Berlin, 2008).

[48] T. Tufarelli, A. Ferraro, M. S. Kim, and S. Bose, *Reconstructing the Quantum State of Oscillator*

Networks with a Single Qubit, Phys. Rev. A **85**, 032334 (2012).
[49] A. N. Cleland (private communication).
[50] A. Megrant et al., Planar Superconducting Resonators with Internal Quality Factors above One Million, Appl. Phys. Lett. **100**, 113510 (2012).
[51] H. Paik et al., Observation of High Coherence in Josephson Junction Qubits Measured in a Three-Dimensional Circuit QED Architecture, Phys. Rev. Lett. **107**, 240501 (2011).
[52] C. Rigetti et al., Superconducting Qubit in a Waveguide Cavity with a Coherence Time Approaching 0.1 ms, Phys. Rev. B **86**, 100506(R) (2012).
[53] R. Barends et al., Coherent Josephson Qubit Suitable for Scalable Quantum Integrated Circuits, arXiv:1304.2322.
[54] J. Bourassa (private communication).
[55] D. L. Underwood, W. E. Shanks, J. Koch, and A. A. Houck, Low-Disorder Microwave Cavity Lattices for Quantum Simulation with Photons, Phys. Rev. A **86**, 023837 (2012).